\documentclass{article}
\usepackage{spconf,amsmath,graphicx, amssymb, svg, booktabs, multirow, subcaption}

\title{AITCHISON GEOMETRY ON THE SIMPLEX FOR UNCERTAINTY QUANTIFICATION IN BAYESIAN HYPERSPECTRAL IMAGE UNMIXING}

\name{Hector Blondel$\textsuperscript{1}$, Lucas Drumetz$\textsuperscript{1,2}$, Thierry Chonavel$\textsuperscript{1}$\thanks{This work was supported by France 2030 and ClusterAI SequoIA through AI Chair GENESIS.}}
\address{$\textsuperscript{1}$ IMT Atlantique, Lab-STICC, UMR CNRS 6285, Plouzané, France\\ $\textsuperscript{2}$ INRIA ODYSSEY team-project, Brest, France}
\begin{document}
%
\maketitle
\begin{abstract}
Most algorithms for hyperspectral image unmixing produce point estimates of fractional abundances of the materials to be separated. However, in the absence of reliable ground truth, the ability to perform abundance uncertainty quantification (UQ) should be an important feature of algorithms, e.g. to evaluate how hard the unmixing problem is and how much the results should be trusted. The usual modeling assumptions in Bayesian unmixing rely heavily on the Euclidean geometry of the simplex and typically disregard spatial information. In addition, to our knowledge, abundance UQ is close to nonexistent in the literature. In this paper, we propose to leverage Aitchison geometry used in compositional data analysis to provide practitioners with alternative tools for modeling prior abundance distributions. In particular, we show how to design simplex-valued Gaussian Process priors using this geometry. Then we link Aitchison geometry to constrained optimization and sampling algorithms, and propose UQ diagnostics that comply with the constraints on abundance vectors. We illustrate these concepts on real and simulated data.
\end{abstract}
\section{Introduction}
\label{sec:intro}

Hyperspectral image unmixing is a classical source separation problem in remote sensing~\cite{bioucas2012hyperspectral}, wherein the spectral signatures of the materials present in the image (endmembers) and their relative proportions in each pixel (abundances) have to be estimated, with numerous applications in e.g. environmental monitoring~\cite{ellis2004evaluation}, urban planning~\cite{marinoni2015accurate}, planetology and astrophysics~\cite{ceamanos2011intercomparison}. In spite of a plethora of algorithms taking into account difficulties such as the nonlinearity of the mixing phenomena~\cite{heylen2014review} or the intra class variability of the endmembers~~\cite{borsoi2021spectral}, the domain is plagued by the difficulty of rigorously validating algorithms on real data, due to the lack of reliable ground truth~\cite{cavalli2023spatial}. Simply assessing the difficulty of a specific instance of the problem performing uncertainty quantification (UQ) in the estimated quantities depends on many factors, ranging from the spatial and spectral resolutions, the granularity of the ``pure" materials to separate, or the observation conditions. Even though a recent trend is to use data-driven probabilistic models to capture the variability of the endmembers (see e.g.~\cite{zhu2025diffusion,shi2021probabilistic}), to our knowledge, the associated uncertainty on the abundances has not been explored or quantified extensively. The reasons for this may be that 1) Abundances are subject to constraints, making it harder to interpret uncertainty. Indeed, typically abundances belong to the unit simplex (though this can be relaxed in several ways) 2) The abundances are a 2D field, i.e. it is desirable and sometimes necessary (for noisy or incomplete data) to account for spatial information, which is not straightforward 3) Handling the posterior distribution is typically intractable and requires sophisticated sampling algorithms, in particular in the presence of constraints~\cite{brosse2017sampling, Hsieh_Kavis_Rolland_Cevher_mirrored_langevin}. Thus, in this paper, we focus on modeling and representing the spatial and material-wise uncertainty of the abundances $p(\mathbf{A})$, where $\mathbf{A} \in \mathbb{R}^{P\times N}$ with $P$ the number of materials and $N$ the number of pixels. Each column $\mathbf{a}_n$ of $\mathbf{A}$ belongs to the probability $(P-1)$-simplex $\Delta^{P-1} = \{\mathbf{a}\in [0,1]^P | \sum_{k=1}^P a_k = 1\}$. We work in an idealized Bayesian linear mixing model framework where we assume an endmember is represented by a single known signature: i.e. the likelihood of the observations $\mathbf{X}\in \mathbb{R}^{L\times N}$ with $L$ the number of spectral bands is given by $p(\mathbf{X}|\mathbf{A}) = \mathcal{N}(\mathbf{SA}, \sigma^2 \mathbf{I}_L)$ with $\mathbf{S} \in \mathbb{R}^{L \times P}$ is the endmember matrix (containing the endmember signatures in its columns). With this setup, we aim to sample and represent the uncertainty of the posterior distribution:
\begin{equation}
    \label{eq:posterior}
    p(\mathbf{A}|\mathbf{X)} \propto p(\mathbf{X}|\mathbf{A})p(\mathbf{A})
\end{equation}
Dirichlet distributions are classically used as a natural model for the prior distribution of abundance vectors~\cite{nascimento2011hyperspectral}. In compositional data analysis~\cite{Aitchison_1982}, however, another geometry, so-called Aitchison geometry, is used because of favorable properties and natural handling of the constraints. To our knowledege, this geometry has not been considered in hyperspectral image unmixing: hence this paper investigates the potential benefits of such a geometry with a focus on a Bayesian view of the problem. We first review the basics of Aitchison geometry in a unmixing context. Then we show how it 1) allows to leverage bijections between $\mathbb{R}^{P-1}$ and the interior of the simplex to define a new geometry that 2) can be used to extend Gaussian Processes (GP) to simplex-valued random fields by \emph{pushing forward} Euclidean GP so that they respect the constraints. The resulting GP make for spatialized geometric priors allowing closed-form interpolation of abundance maps and provide sensible priors for spatialized Bayesian unmixing. 3) may be leveraged in optimization or sampling algorithms. 4) provide principled tools to define constraint-aware ways to represent and visualize abundance uncertainty at the pixel and image levels.\\
\vspace{-0.6cm}

\section{Aitchison Geometry}

\label{sec:prior}

\begin{figure}
    \centering
     \includegraphics[width=0.7\linewidth]{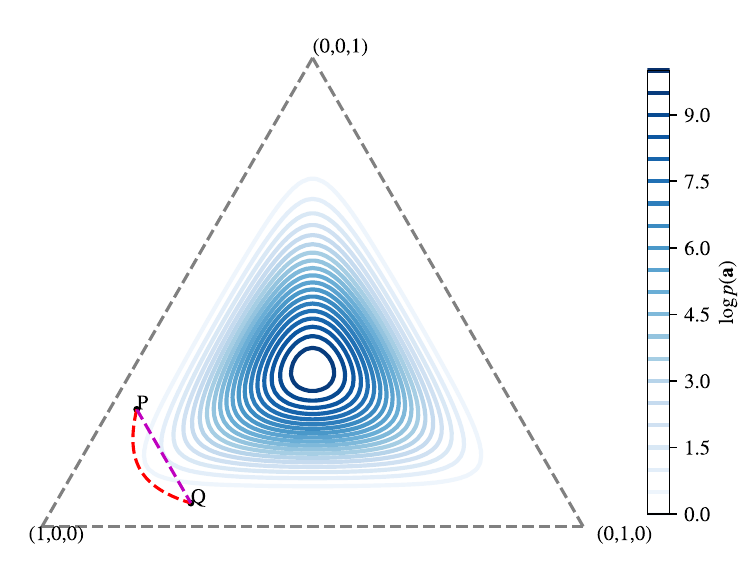}
    \caption{ilr-gaussian pdf on the 3-simplex : $\textrm{ilr}(\mathbf{a})\sim \mathcal{N}(\mathbf{0},\sigma_a^2 \mathbf{I}_2)$, $\sigma_a^2 = 0.25$ and geodesics between two points under the Euclidean (purple) and Aitchison metrics (red).}
    \label{fig:prior}
\end{figure}

In this section, we recall how diffeomorphic transformations $\psi : \textrm{int}\ \Delta^{P-1} \to \mathbb{R}^{P-1}$ from compositional data analysis turn the interior of the simplex into a flat Riemannian manifold with a pullback Euclidean metric. This allows to design prior distributions on the simplex that are pushforwards of Gaussian distributions on $\mathbb{R}^{P-1}$.\\
A straightforward choice for such a $\psi$ is the alr transformation \cite{Aitchison_1982} $\textrm{alr}(\mathbf{a}) := [\log a_k - \log a_P]_{1\leq k \leq P-1}$, which removes one coordinate of $\mathbf{a}$ to get rid of the sum-to-one constraint, and then takes the logarithm to alleviate the positivity constraint. However, the choice of the coordinate to remove is arbitrary, introducing symmetry-breaking artifacts in the representation. A way to symmetrize the previous transformation is the Centered Log Ratio (clr)~\cite{Aitchison_1982}:
\begin{equation}
    \mathbf{w} = \psi(\mathbf{a}) \triangleq \textrm{clr}(\mathbf{a}) := \begin{bmatrix}
    \log a_p - \frac{1}{P} \sum_{k=1}^P \log a_k \\
    \end{bmatrix}_{p = 1,...,P}
\end{equation}
$\textrm{clr}: \textrm{int} \ \Delta^{P-1} \to \mathbb{R}^{P-1}$ is not a bijection, but the restriction onto its image, the hyperplane $H$ with normal vector $\mathbf{1}\in \mathbb{R}^{P}$, is a diffeomorphism. For numerical computations, choosing an orthonormal basis $\mathbf{H}\in \mathbb{R}^{{P}\times{P-1}}$ of $H$ allows to work in $\mathbb{R}^{P-1}$, and the resulting transformation is called the Isometric Log Ratio (ilr) $\mathbf{z} = \textrm{ilr}(\mathbf{a}) \triangleq \mathbf{H}^T\mathbf{w}$. The inverse transformation is the \textrm{softmax} function:
\begin{equation}
    \textrm{ilr}^{-1}(\mathbf{z}) = \textrm{softmax} (\mathbf{H}\mathbf{z}) = \Bigg[ \frac{\exp{w}_p}{\sum_{k=1}^{P} \exp(w_k)} \Bigg]_{p=1,...,P}. 
\end{equation}
With a slight abuse of notation, we still refer to $\textrm{ilr}(\mathbf{a})$ as $\psi(\mathbf{a})$. We note that $\psi$ (as any diffeomorphism $\textrm{int} \ \Delta^{P-1} \to \mathbb{R}^{P-1}$) turns $\textrm{int} \ \Delta^{P-1}$ into a flat Riemannian manifold with at metric that is the pullback $\psi^*g$ of the Euclidean metric $g$ on $\mathbb{R}^{P-1}$. In other words, $(\textrm{int} \ \Delta^{P-1}, \psi^*g)$  is isometric to $(\mathbb{R}^{P-1},g)$ by construction. The associated geodesic distance is given by the Euclidean distance in $\textrm{ilr}$ space:
\begin{equation}
    d_\psi(\mathbf{a}, \mathbf{a}') = ||\textrm{ilr}(\mathbf{a})-\textrm{ilr}(\mathbf{a}')||_2
\end{equation}
In particular, geodesics are obtained in the simplex by applying $\textrm{ilr}^{-1}$ to straight lines in $\mathbb{R}^{P-1}$. This is a crucial difference from the usual Euclidean metric on $\Delta^{P-1}$ that is induced from the Euclidean metric on $\mathbb{R}^P$. 
For the Euclidean metric, the geodesics are straight lines in the simplex. To illustrate the fundamental differences between both metrics, see Fig.~\ref{fig:prior}, where we plot both geodesics to joint two points of the simplex (P and Q) that correspond to the same abundances, up to a permutation of two of their components, and the other fixed. The Euclidean geodesic (in purple) is a straight line, and all abundances on the way will retain the same fraction of material $(1,0,0)$, mixing linearly the two remaining materials. By constrast, the Aitchison geodesic (in red) first removes the excess of material $(0,1,0)$ (getting closer to the vertex corresponding to pure material $(1,0,0)$), before gradually adding the missing contribution of material $(0,0,1)$. In fact, the interpolation is linear in ilr space.\\
We can define an uninformative prior for abundance vectors using ilr. A minimal requirement for such prior densities $p(\mathbf{a})$ is to be invariant by permutation of the coordinates whenever the distribution on $\mathbf{z} = \textrm{ilr}(\mathbf{a})$ is zero mean and rotation invariant.
We will call this property ``isotropy" hereafter. The most natural is to consider isotropic Gaussian latent distributions
\begin{equation}
    \psi(\mathbf{a}) = \mathbf{z} \sim \mathcal{N}(\mathbf{0}, \sigma_a^2 \mathbf{I}_{P-1})
\end{equation}
The resulting prior in the simplex can be easily shown to be isotropic using the change of variables formula for probability densities. Indeed, the pdf writes in the simplex :
\begin{equation}
    p(\mathbf{a}) \propto \frac{1}{\prod_{k=1}^P a_k } \exp \left( - \frac{1}{2 \sigma_a^2} \| \psi(\mathbf{a}) \|_2^2\right),
    \label{eq:pdf}
\end{equation}
and is unchanged when permuting the components of vector $\mathbf{a}$. A visual representation of this density is given in Figure~\ref{fig:prior}, for $\sigma_a^2 = 0.25$. Note that the distribution becomes multimodal (but remains permutation invariant) for large enough values of $\sigma_a^2$. Shifting the mean of the Gaussian latent can bias the prior towards a vertex or an edge, while arbitrary covariance matrices lead to more intricate behavior, providing a rich class of priors.
While not obvious from Eq.~\eqref{eq:pdf} due to the Jacobian term, the density goes to 0 when $\mathbf{a}$ tends to the boundary of the simplex: the prior forbids abundances with strictly zero components. In numerical computations, we can get sufficiently close to the boundary so that this does not matter in practice.
Finally, let us point out that Dirichlet priors (and in particular a uniform prior), which are intrinsically tied to the Euclidean geometry, do put nonzero mass to the boundary. A consequence is that the posterior distribution for a Gaussian likelihood will be a truncated Gaussian on the simplex, which requires renormalization and does not correspond to physically realistic abundance variations. By contrast, posterior distributions with an ilr-Gaussian prior will naturally approach the simplex boundary due to the log-barrier.

\section{Simplex-valued Gaussian processes}
\label{sec:GP}
\subsection{Spatializing a pushforward Gaussian prior}

In this section, we use the ilr transformation to design a simplex-valued spatialized prior on the abundance image using latent GP, which we call pushforward GP.\\
For a spatial process with values on the simplex $\mathbf{a}(\mathbf{u)}$, where $ \mathbf{u} = (x,y) \in \mathbb{R}^2$ are the spatial coordinates of a pixel, we want $\mathbf{u} \mapsto \psi(\mathbf{a}(\mathbf{u}))$ to follow a vector-valued GP~\cite{alvarez2012kernels}. Given $N$ points $\mathbf{u}_1, \dots, \mathbf{u}_N$,  
the vectorized version of latent space coefficients $\mathbf{Z} = \psi(\mathbf{A}) \triangleq [\psi(\mathbf{u}_1), \dots, \psi(\mathbf{u}_N)] \in \mathbb{R}^{(P-1)\times N}$ is a multivariate Gaussian with a $(P-1)N \times (P-1)N$ block covariance matrix 
\begin{equation}
    \mathbf{K}_Z =  \begin{bmatrix}
 \mathbf{K}(\mathbf{u}_1, \mathbf{u}_1)& \dots & \mathbf{K}(\mathbf{u}_1, \mathbf{u}_N) \\
 \vdots &  & \vdots \\
 \mathbf{K}(\mathbf{u}_N, \mathbf{u}_1)& \dots & \mathbf{K}(\mathbf{u}_N, \mathbf{u}_N)  \\
\end{bmatrix}.
\end{equation}
Each $\mathbf{K}(\mathbf{u}_i, \mathbf{u}_j)\in \mathbb{R}^{(P-1) \times (P-1)}$ is a positive definite matrix. To be able to fix the global covariance structure of the GP in an intuitive and tractable way, we make the \textit{separability assumption}, by expressing the overall kernel matrix as a Kronecker product  $\mathbf{K}_Z = \mathbf{K}_P \otimes \mathbf{K}_U$ where $\mathbf{K}_U \in \mathbb{R}^{N \times N}$ is the Gram matrix coming from the discretization of a real-valued spatial kernel, and $\mathbf{K}_P \in \mathbb{R} ^{(P-1)\times (P-1)}$ is a positive definite matrix encoding correlations between entries of the latent vector (see \cite{alvarez2012kernels}). In the absence of further information, the latent space representation is made isotropic in every pixel: $\mathbf{K}_P =\sigma_a^2\mathbf{I}_{P-1}$. The latent space coefficients $\psi(\mathbf{A})$ then follow a centered matrix normal distribution: 
\begin{equation}
  \psi(\mathbf{A}) \sim \mathcal{MN}(\mathbf{0}_{(P-1)\times N}, \sigma_a^2\mathbf{I}_{P-1}, \sigma_k^2\mathbf{K_U}). 
\end{equation}




\noindent $\sigma_k^2$ (fixed to 1 throughout our experiments) scales the spatial kernel in latent space. For $\mathbf{K}_U$, we use the exponential kernel, with characteristic length-scale $l$:
$k(\mathbf{u},\mathbf{u}') =   \exp \left(-\frac{1}{l}\left \|\mathbf{u} - \mathbf{u'}\right \|_2\right)$.
This kernel (among other edge preserving choices that are out of scope of this paper) is better suited (and numerically more stable~\cite{stein1999interpolation}) than the Gaussian kernel for spectral unmixing, where abundance maps present discontinuities. 
With the change of variable formula, the pdf of $\mathbf{A}$ is:
\begin{multline}
p(\mathbf{A}) = \left| \det \nabla \psi(\mathbf{A}) \right| p_{\psi(a)}(\psi(\mathbf{A})) \\
\propto \frac{1}{\prod_{l=1}^N \prod_{k=1}^P a_{kl}}  \exp \left( - \frac{1}{2} \Big\| \psi(\mathbf{A}) \mathbf{K_U}^{-\frac{1}{2}} \Big\|_F^2 \right)
\end{multline}
where $\| . \|_F$ is the Frobenius norm.

\subsection{Interpolating partial and noisy abundance maps}
As in the unconstrained case, the pushforward GP allows us to interpolate partially observed abundance maps (e.g. in the presence of clouds or from discrete in situ measurements as in the geostatistics literature~\cite{clarotto2022new}) in closed form.
Without going into details, this simply amounts to fitting a Euclidean GP after moving the abundances to $\textrm{ilr}$ space, before moving back to abundance space using the $\textrm{softmax}$ function. In other words, conjugacy properties of the Gaussian distribution in latent space carry over to simplex-valued interpolation. 

\subsection{Sampling the unmixing posterior: Mirror Langevin}
\label{subsec:mirrored-langevin}


Our pushforward GP can also be used as a spatial prior for the abundance maps in Bayesian spectral unmixing, as in Eq.~\eqref{eq:posterior}. As we will see, the problem is more challenging than interpolating spatialized abundances, since now the observations do not live on the simplex, due to the mixture induced by the endmember matrix $\mathbf{S}$. As a result, the posterior expression will be intractable in general: the observation likelihood $p(\mathbf{X}|\mathbf{S},\mathbf{A})$ is (matrix) Gaussian distribution in the Euclidean space $\mathbb{R}^{L\times N}$, but the prior is (matrix) Gaussian in the \emph{latent} $\textrm{ilr}$ space, breaking the conjugacy properties of classical GP regression:
\begin{multline}
    p(\mathbf{A}|\mathbf{X},\mathbf{S}) 
     \propto \frac{1}{\prod_{l=1}^N \prod_{k=1}^P a_{kl}}  \exp \left( - \frac{1}{2} \Big\| \psi(\mathbf{A}) \mathbf{K_U}^{-\frac{1}{2}} \Big\|_F^2 \right) \\   \exp \left( - \frac{1}{2\sigma^2} \Big\|  \mathbf{X} - \mathbf{S}\mathbf{A} \Big\|_F^2 \right)
\end{multline}


We must then resort to sampling algorithms (or variational inference) in order to manipulate the posterior, compute estimators and perform UQ. A well-known sampling technique among Markov Chain Monte Carlo (MCMC) methods is the Unajusted Langevin Algorithm (ULA), which can be seen as the sampling counterpart to the MAP point estimate that gradient descent on the negative log posterior targets. However, as our search space (the simplex) is constrained, the latter method cannot be used as is. ``Euclidean fixes" are projected gradient/Langevin algorithms that use the Euclidean projection on the simplex~\cite{condat2016fast}. Alternatively, when provided with a convex potential $\varphi$ whose gradient is a bijection matching a constrained space that is a convex subset of Euclidean space to a new latent Euclidean space, the mirror descent algorithm~\cite{yudin1983problem} allows to handle the constraint seamlessly in this new domain. Similarly to ULA, a sampling version of mirror-descent, so-called ``Mirror Langevin" was first proposed in \cite{Hsieh_Kavis_Rolland_Cevher_mirrored_langevin}, and adapts the Mirror Descent algorithm to sample in a constrained space thanks to the mirror map $\nabla \varphi$. We refer to \cite{Hsieh_Kavis_Rolland_Cevher_mirrored_langevin} for details on the update rules. 

Mirror Langevin can then be seen as a kind of ULA in a different geometry induced by the mirror map. In the original mirror Langevin paper \cite{Hsieh_Kavis_Rolland_Cevher_mirrored_langevin}, $\textrm{alr}$ is shown to be a valid mirror map and is proposed for sampling on the simplex. Instead, we recommend to use the previously defined $\textrm{ilr}$ transformation we used for defining our isotropic prior, as it is also the gradient $\textrm{ilr} = \nabla h$ of the scalar convex potential function (the entropy) $h : \mathbf{a} \in \Delta^{P-1} \mapsto \sum_{k=1}^P a_k \log a_k$. 
 This choice is again motivated by the isotropy of the $\textrm{ilr}$ function and also allows to use the same underlying geometry for sampling as the one used for the prior (contrary to, say, a projected ULA algorithm), which may be favorable for convergence. We use this algorithm to obtain posterior samples hereafter.


\section{Representing Uncertainty: confidence regions and total variance}
\label{sec:UQ}


In this section, we show how to perform UQ on the abundances in practice for the unmixing problem, given samples from the posterior from the previous section. First, we deal with defining the posterior mean and uncertainty measures, and visualizing them (up to $P=4$, which covers many cases of practical interest, given that that more than 4 materials significantly contribute to a pixel's signature is rare) at the level of a single pixel. Then we show how to visualize the posterior mean and different notions of variance at the image level. After sampling, in a classical Euclidean space, the usual pointwise estimator for the parameters is given by the posterior Euclidean mean as it is the minimum mean square estimator of $\mathbf{A}$ given $\mathbf{X}$. The pullback Euclidean geometry defined by ilr suggests defining the mean in the latent space \cite{Pawlowsky-Glahn_Egozcue_2001}, which corresponds to the geodesic mean (which is in fact the minimum squared geodesic distance estimator). For $M$ samples $\mathbf{A}^1, \dots, \mathbf{A}^M$ of a simplex-valued image the geodesic mean in pixel $n$, $\bar{\mathbf{a}}_n$, is the softmax of the Euclidean mean in ilr space: 
\begin{equation}
    \bar{\mathbf{a}}_n \triangleq \psi^{-1} \left( \frac{1}{M}\sum_{m=1}^M \psi(\mathbf{a}_n^m) \right )
    \vspace{-0.3cm}
\end{equation}
\subsection{Pixelwise UQ: confidence regions in the simplex}
\label{subsec:pixel_level}
\begin{figure}
    \centering    
    \includegraphics[width=0.35\textwidth]{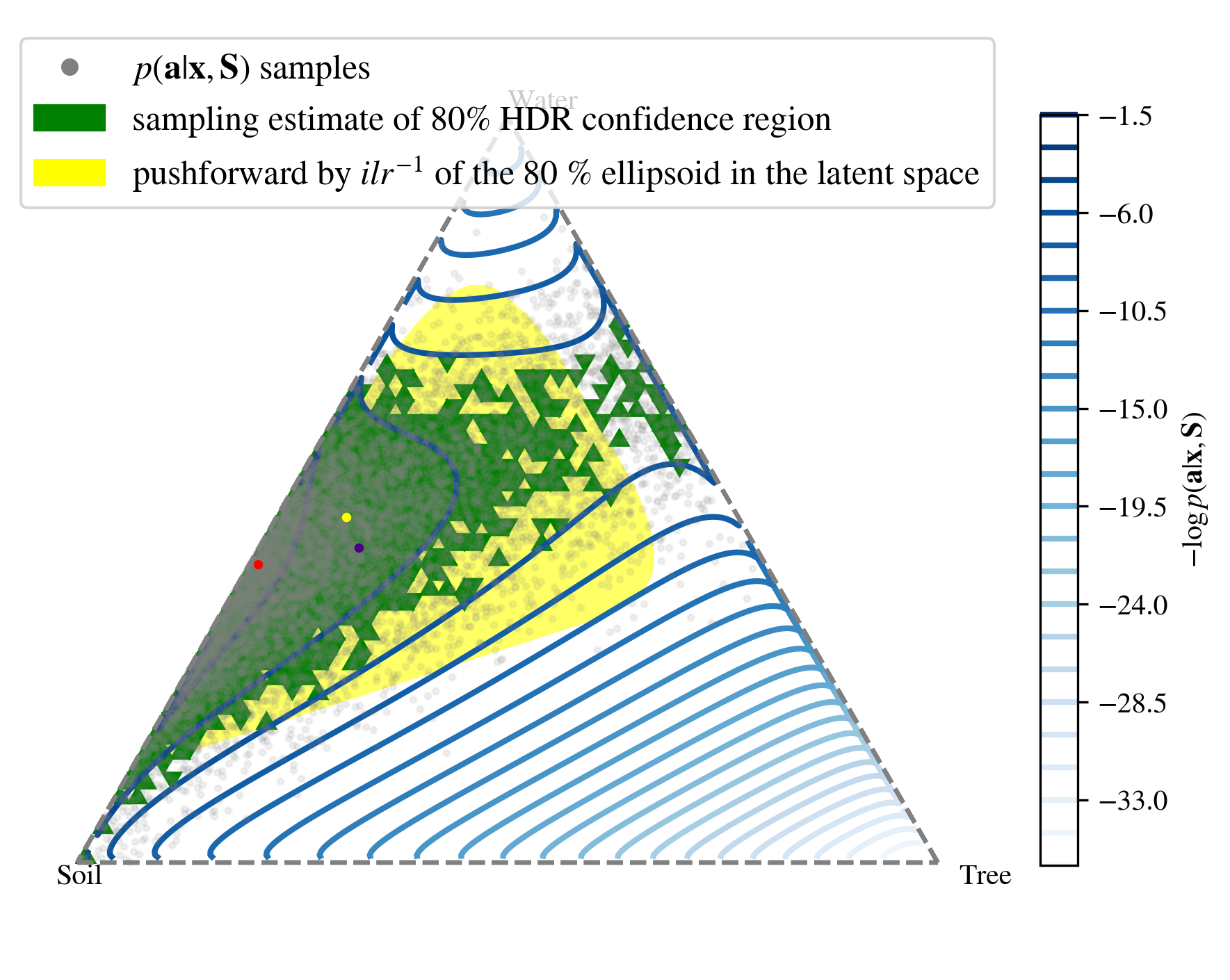}
    \caption{Unmixing abundances $[0.59, 0.01, 0.4]$ (red), $\text{SNR} = 8 ~\text{dB}$ with a multimodal posterior, sampled 10000 times. Yellow dot : geodesic mean $\bar{\mathbf{a}}$ ; purple dot : Euclidean mean.}
    \label{fig:samson10dB}    
\end{figure}

As the posterior distribution is no longer Gaussian (neither in latent nor simplex space), the marginal densities on individual pixels $\mathbf{a}_n$ are not known in closed form due to the spatial correlations introduced by the GP prior. Hence, samples are necessary to visualize the posterior. A naive way to define confidence regions would be to make a Gaussian approximation of the posterior, using the classical covariance, and reporting confidence ellipsoids. However, in addition to poor approximations for complex posteriors, the regions may not be entirely in the simplex (even if all samples are), which makes the geometric interpretation meaningless. A geometry compliant strategy would be to define a Gaussian approximation of the posterior in latent space, so that latent confidence ellipses may be transformed back using the softmax function. Still, the target distribution might be complex or even multimodal in latent space, making the approximation poor, whilst capturing these effects is important in UQ.\\
For visualizing uncertainty, we propose a method for estimating confidence regions on individual pixels. A confidence region of threshold $\alpha$ for the the random variable $\mathbf{a}|\mathbf{X}$ is a region $\mathcal{I}_\alpha \subset \Delta^{P-1}$ such that $\mathbb{P}(\mathbf{a}\in I_{\alpha} | \mathbf{X}) \geq 1 - \alpha$. Such regions are not unique in general, so we opt for the minimum volume region, which can be proven to be the set of all points for which the density is greater than a certain threshold~\cite{hyndman_computing_1996}, called \textit{highest density region} (HDR). By construction, this set, for a suitable $\alpha$ reveals the presence of different modes through several connected regions. For any $\mathbf{a}$, a convergent estimator of the HDR can be found as follows: we estimate the density via the $M$ samples $\mathbf{a}^1, \dots, \mathbf{a}^M$ through a histogram in the simplex (or any density estimator). Then, we sort them by increasing order of their densities ($p(\mathbf{a} = \mathbf{a}^1) \leq \dots \leq p(\mathbf{a} =\mathbf{a}^M)$), and take the $\lfloor \alpha M \rfloor$-s smallest, $p(\mathbf{a}^{\lfloor \alpha M \rfloor})$, where $\lfloor  \cdot \rfloor$ is the floor function.\\
We illustrate the usefulness of this construction by simulating data using three endmembers extracted from the Samson dataset (see Section~\ref{subsec:image_level} for a description): soil, vegetation and water and respective GT abundances $[0.59, 0.01, 0.4]$. We model the abundances from a multimodal isotropic ilr-Gaussian prior ($\sigma_a^2 = 5$) (this is not a realistic modeling choice but results in a multimodal posterior for illustrative purposes), and add Gaussian noise (SNR = 8dB) to the mixture. Fig.~\ref{fig:samson10dB} shows the posterior distribution and corresponding samples.
The posterior distribution tends to incorporate vegetation in the mixture due to noise and a badly specified prior, and presents two modes and a plateau of likely abundances under the posterior. First, we note that a large enough Gaussian confidence ellipse would leave the simplex. The pullback of a latent confidence ellispe (while contained in the simplex) is not able to reach edges of the simplex, which have high density. We see that the HDR region captures the multimodal nature of the distribution by construction. 
\vspace{-0.2cm}
\subsection{UQ at the image level: geodesic mean and variance}
\label{subsec:image_level}
We now propose ways to represent the uncertainty at the image level, using a single scalar value representing the total variability in each pixel.
Total variance (the trace of the Euclidean covariance matrix for the marginal distribution) is classically used as a scalar estimate of the variability around the mean in every pixel. As noted above, the Euclidean covariance matrix is not directly interpretable due to the constraints. In particular, the diagonal elements measure variances along the coordinate axes, ignoring that the simplex is a convex subset of a hyperplane inside $\mathbb{R}^P$.
In ilr geometry, the sample geodesic total variance in pixel $n$, $\textrm{TV}_n$ is the classical total variance in latent space (i.e. the trace of the empirical latent covariance matrix), and corresponds to the mean squared geodesic distance to $\bar{\mathbf{a}}_n$: 
\begin{equation}
    \textrm{TV}_n \triangleq \frac{1}{M-1}\sum_{m=1}^M || \psi(\mathbf{a}^m_n) -\psi(\bar{\mathbf{a}}_n) ||_2 ^2
\end{equation}
Componentwise variances at the pixel level (diagonal elements of the Euclidean covariance matrix) are used as marginal measures of variability. These can become misleading for abundance vectors, due to the sum-to-one constraint. However, these quantities make sense in ilr space, and reflect the dispersion in the log-ratio between pairwise abundances.

\begin{figure}
       
    \centering    
 \includegraphics[width=0.39\textwidth]{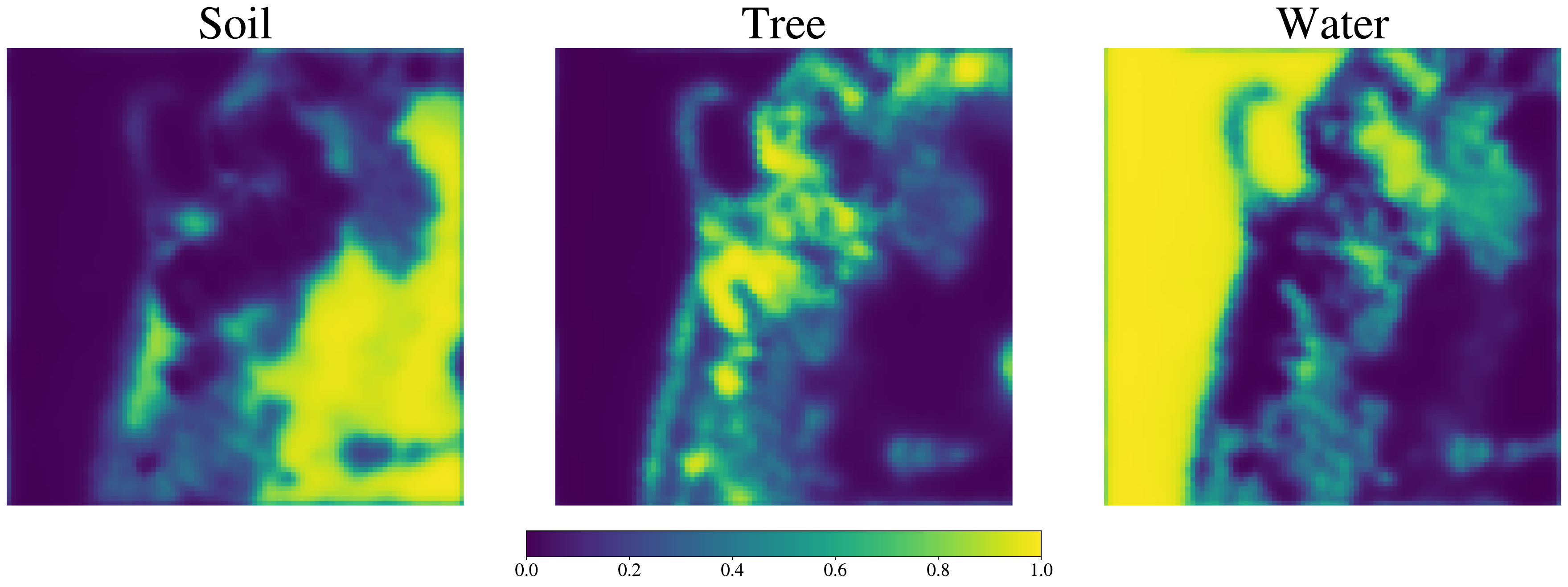}
    \caption{Geodesic abundance maps for a pushforward GP prior.}
    \label{fig:means}    
\end{figure}
\begin{figure}[t]
    \begin{subfigure}{0.48\textwidth}
        \centering  
\includegraphics[width=0.9\textwidth]{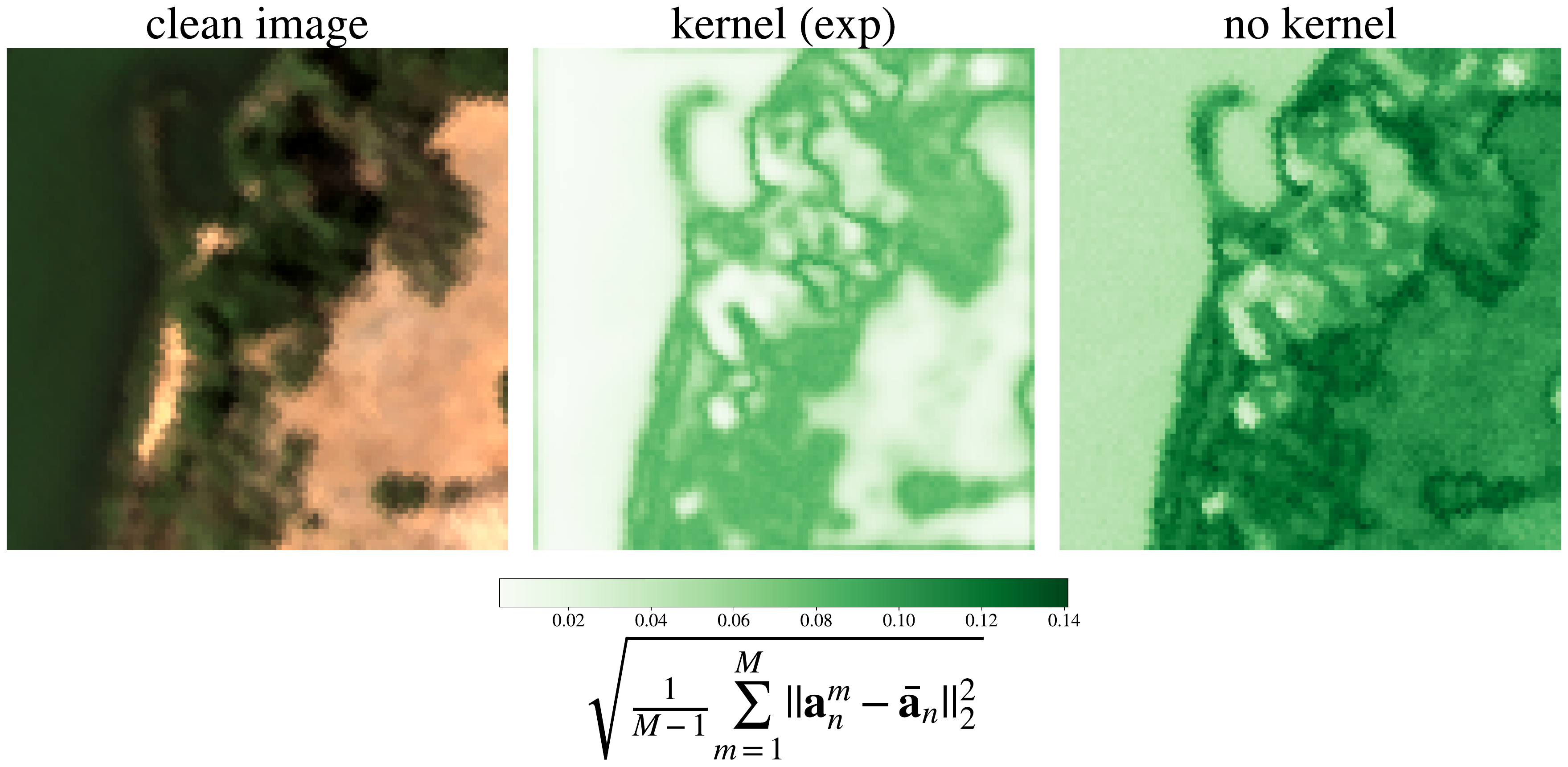}        
    \end{subfigure}
     \begin{subfigure}{0.48\textwidth}
        \centering    
                \includegraphics[width=0.9\textwidth]{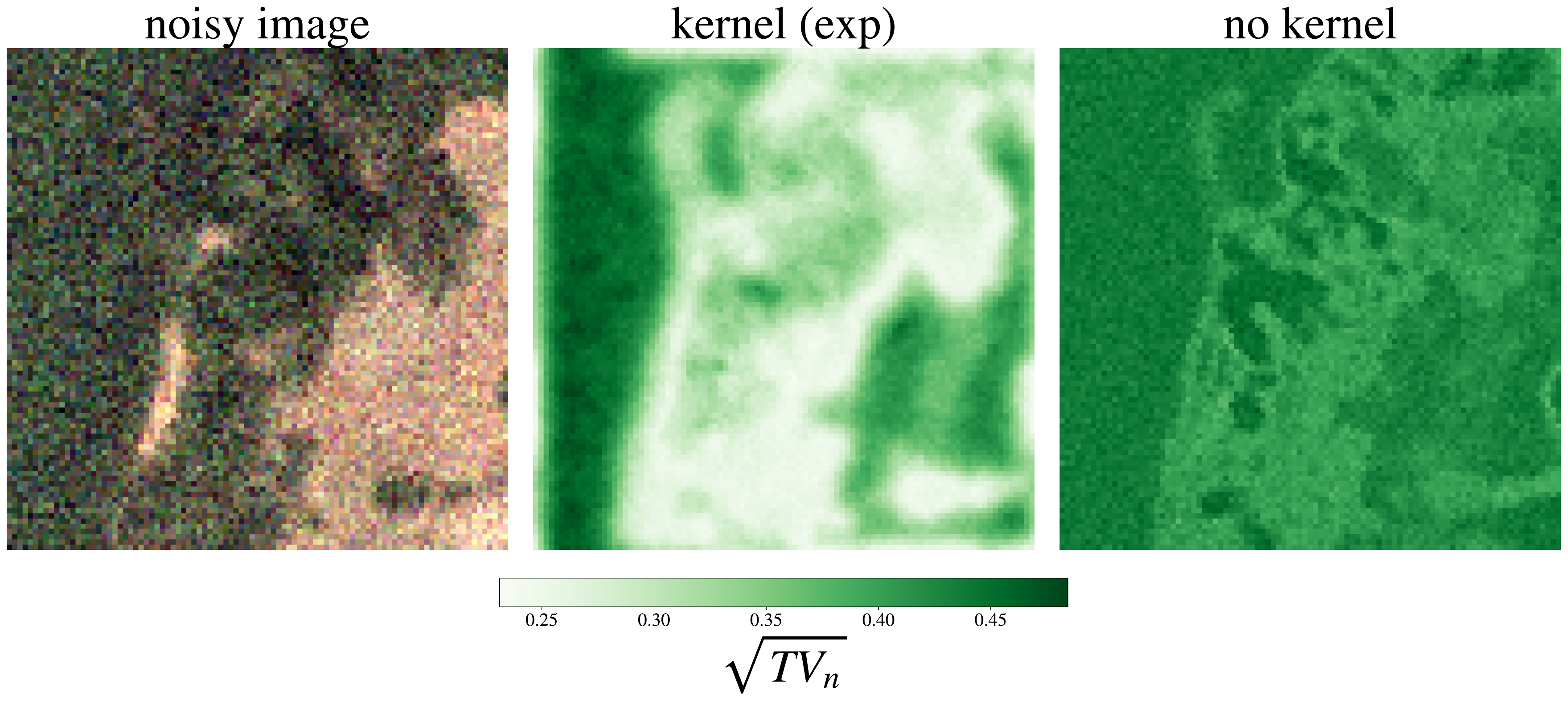}
    \end{subfigure}
    \title{standard deviation for samples of bayesian posterior model with and without kernel}
    \vspace{-0.6cm}
\caption{Euclidean (top) and Aitchison (bottom) standard deviations maps for spatial and non spatial priors.}
\label{fig:stds}  
\vspace{-0.3cm}
\end{figure}

We illustrate these notions on the Samson dataset. It consists in  $95 \times 95$ pixels containing the reflectance measured on $L=195$ wavelengths spreading from 401 nm to 889 nm. A RGB composition is shown in Fig.~\ref{fig:stds}. We have artificially added $15 \text{dB}$ noise to highlight differences between kernel-based and pixelwise algorithms. We use same 3 endmembers as in Section \ref{subsec:pixel_level}. We compare a pushforward GP prior, with $l=10$ pixels and $\sigma_a^2 = 0.25$ and a non-spatialized prior (using a dirac kernel in space), drawing 1000 samples. The geodesic mean abundance maps for the GP prior are shown in Fig. \ref{fig:means} (the Euclidean means are visually similar and are not shown).
Geodesic and Euclidean total standard deviations are shown in Fig. \ref{fig:stds} for both models.
The first and main observation is that Euclidean variance is low in close to pure areas and high in more mixed regions, while Aitchison variance shows opposite behavior. The noise in the observations causes small absolute fluctuations in each component: this is reflected by a relatively low total standard deviation. These small variations have a larger impact on the relative importance between the materials (the ratios between components), which explains the larger Aitchison variance. This behavior is expected near the boundaries of the simplex, where Aitchison geometry is more sensitive due to the log barrier. On the other hand, in more mixed areas, larger absolute variations in the components reflect in Euclidean variance, but tend to change the ratios between materials less (the relative importance of the materials changes less), leading to smaller Aitchison variance. Both notions of variance bring complementary information.
We also see the interest of having a spatial prior: beyond just obtaining smoother abundance maps, uncertainty is also smoother (it should not be independent from one pixel to the next), and is comparatively smaller in the kernel version of the models, eliminating the perturbations due to noise.

\vspace{-0.15cm}
\section{Conclusion}
\label{sec:ccl}

In this paper, we have presented how to endow the simplex with a non Euclidean geometry that complements usual tools for spectral unmixing implicity 
on Euclidean geometry. We have shown in particular how this geometry allows to define simplex-valued Gaussian Processes to design spatialized priors in Bayesian models. We have also shown the potential of this geometry to perform uncertainty quantification in unmixing, both at the pixel level and for the whole image. Future work will include exploring other UQ diagnostics and using Aitchison geometry to define probabilistic evaluation metrics to validate Bayesian algorithms.

\vfill\pagebreak

\bibliographystyle{IEEEbib}

\end{document}